# Pair correlation and dynamic Jahn-Teller effect: high $T_c$ in nanoclusters

Vladimir Kresin[(1)], Yurii Ovchinnikov[(2)], Jacques Friedel[(3),*]

[(1)]*Lawrence Berkeley Laboratory, University of California at Berkeley, CA 94720 USA*
[(2)]*,L.Landau Institute for Theoretical Physics, RAN, Moscow, 117334 Russia*
[(3)]*Laboratoire de Physique des Solids, Universite Paris XI, Batiment 510-91405 Orsay Cedex, France, EU*

Abstract

Electronic states in metallic nanoclusters form energy shells and degree of their filling depends on the number of delocalized electrons. In the region close to half-filling the cluster's geometry oscillates between the prolate and oblate configurations (dynamic Jahn-Teller effect). For large clusters ($N>10^2$ ; $N$ is the number of delocalized electrons) this effect competes with pair correlation and, as a result, it is perfectly realistic to observe the transition to the superconducting state. For some clusters (*e.g.*, for $Zn_{76}$, $Al_{70}$) the value of the critical temperature is rather high ( $\gtrsim 140K$).

* deceased

This Letter is concerned with the superconducting state of metallic nanoclusters. It is a continuation of our study of dynamic Jahn-Teller (JT) effect in such nanoparticles [1]. As is known, the delocalized electrons in many metallic nanoclusters form the energy shells similar to those in atoms and nuclei, see, *e. g.*, the reviews [2,3]. Because of it, such clusters are called "artificial" atoms. The shell structure was discovered in alkali clusters [4], and later in clusters of other metals ,e.g., in *Al, Ga, Zn, In* [5-12]. Clusters with complete energy shells form a special group, so-called "magic" clusters (by analogy with "magic" nuclei, which are characterized by complete shells for nucleons, see, e.g., [3,13]). The "magic" clusters with complete shells are very stable similarly to inert atoms. The existence of "magic" numbers with $N=N_m$ *(e.g., $N_m$=20,40,$\cdots$.,138,198, N* is a number of delocalized electrons) is the most profound manifestation of the shell structure. Thanks to the stability, the "magic" numbers can be determined by a number of experimental methods (time-of flight mass spectroscopy [4,9], ionization [7], photoelectron spectroscopy [6.14.15] . As was noted above, the "magic" clusters contain completely occupied energy shells. However, if the highest occupied shell is not completely, but partially filled, we are dealing with the Jahn-Teller (JT) distortion. More specifically, the clusters with slightly occupied shells acquire the prolate structure (see,*e.g.*,[2]) and such a deformation is caused by the usual static JT effect. The same mechanism leads to the oblate configuration for the cluster with almost complete energy shell.

In the region close to half-shell filling we are dealing with two almost degenerate, prolate and oblate, configurations. Such a situation is very favorable for the dynamic JT effect (sometimes it is called "tunneling splitting", see, *e.g.*,[16]). As a result, the dynamic JT effect becomes dominant in the region close to half-filling; this is discussed in detail in our papers [1,17].Then the cluster shape oscillates between the prolate and oblate structures. In the stationary picture the system is described by superposition of these two configurations.

In this paper we focus on the interplay between the dynamic JT effect and electron-vibrational pairing interaction for clusters, which are near half-shell filling . These factors can compete and we demonstrate in this paper that the case when the pairing prevails is perfectly realistic. For such a cluster it is energetically more favorable to make transition into spherical configuration with pairing correlation. This means that clusters with near half-shell filling should display superconducting state, potentially at high temperatures (see below).

The superconducting state of metallic nanoclusters has been studied in a number of papers. Initially such a possibility was considered by Knight [18] and especially by Friedel [19]. The detailed analysis based on the presence of the shell structure with a special focus on "magic" or near "magic" clusters was carried out in our papers [20,21] and later by other authors in [22-24]. The picture of pairing is similar to a well-known concept in nuclear physics [13,26] , where the nucleons (protons, neutrons) form Cooper pairs, and this is a manifestation of the existence of corresponding shell structure.   As we know, Cooper pairs in usual superconductors are formed by electrons with opposite momenta [ 25] (conjugated states).For the clusters the conjugated paired states are formed by electrons with opposite values of projection of orbital momenta (*m,-m*) on the symmetry axis. Pairing in clusters is due to electron-vibrational interaction, so that the mechanism is similar to that in usual superconducting metals.

The equation for the pairing order parameter $\Delta(\omega_n)$ can be written in the following general form [21]:

$$\Delta(\omega_n)Z = \eta \frac{T}{2V} \sum_{n'} \sum_s D(\omega_n - \omega_{n'}, \tilde{\Omega}) F_s^+(\omega_{n'}) \qquad (1)$$

We employ the thermodynamic Green's functions formalism, see, e.g. [27, 28]), and, according to this formalism $\omega_n = (2n + 1)\pi T, n = 0, \pm 1, \pm 2, \ldots$. In Eq.(1) $\Delta(\omega_n)$ is the pairing gap parameter, $F = \Delta(\omega_n)[\omega_n^2 + \xi_s^2 + \Delta^2(\omega_n)]^{-1}$ is the Gor'kov pairing function, $\xi_s = E_s - \mu$ is the energy of the s$^{th}$ electronic state referred to the chemical potential $\mu$, $\eta = \frac{<I>^2}{M\tilde{\Omega}^2}$, $<I>$ is the average electron-ion matrix element, M is the ionic mass, V is the cluster volume, Z is the renormalization function describing an usual electron-ion scattering, and $D = \tilde{\Omega}^2[(\omega_n - \omega_{n'})^2 + \tilde{\Omega}^2]^{-1}$ is the vibrational propagator, $\tilde{\Omega}$ is the characteristic vibrational frequency. Eq.(1) is similar to that in the theory of strong coupling superconductivity ([29], see, *e.g.*, [27,30]),but it contains a summation over discrete energy levels. Note also the presense of the propagatot D makes Eq.(1) different from that in the BCS model,which is valid in a weak coupling approximation .Then $T_c \ll \Omega$;we want to go beyond this restriction.

In addition,contrary to the bulk state (then $\mu=E_F$), the position and temperature dependence of $\mu \equiv \mu(T)$ are determined from the equation reflecting a fixed number of electrons (see, *e.g.*, [27]):

$$N = \sum_{\omega_n} \sum_{S} G_s(\omega_n) e^{i\omega_n \tau} \big|_{\tau \to 0},$$ G is the thermodynamic Green's function. This expression can be reduced to the form:

$$N = \sum_s (u_s^2 \varphi_s^- + v_s^2 \varphi_s^+) \qquad (2)$$

$$u_s^2, v_s^2 = 0.5 \left(1 \mp \frac{\xi_s}{\varepsilon_s}\right), \varphi_s^{\mp} = [1 + \exp{(\mp \varepsilon_s / T)}]$$

For example, for the "magic" cluster, the chemical potential $\mu$ at *T=0K* is located near the middle of the HOS-LUS spacing ( see [21], HOS and LUS are the highest occupied and lowers unoccupied shells, correspondingly; HOS-LUS spacing is similar to HOMO-LUMO spacing in molecular spectroscopy). For clusters with slightly incomplete shells, the chemical potential is located near the HOS edge. As a whole, one must stress that the position of the chemical potential is a factor of key importance for an analysis of various properties of nanoclusters, including the evaluation of $T_c$. The value of $T_c$ can be evaluated from the set of Eqs.(1),(2).

We discussed above two scenarios: dynamic JT effect and the pairing. Let us now consider a cluster with almost half-filled energy shell. Assume, at first, that such a cluster has a spherical shape. Such configuration, however, is unstable, because of orbital degeneracy, caused by the presence of vacant states. This instability can be removed by the dynamic JT effect (see [1]). Then the cluster undergoes the structural change and becomes a superposition of prolate and oblate configurations.

However, there is an another channel. Namely, the electron-vibrational interaction can provide the pairing, and, correspondingly, the transition into superconducting state. Indeed, let us analyze the cluster of interest with use of Eqs.(1) and (2).

Let us introduce the quantities:

$\Phi_n = \Delta(\omega_n)\tilde{\Omega}^{-1}$; $\tau_c = 2\pi T_c \tilde{\Omega}^{-1}$. Then Eq.(1) at $T = T_c$ can be written in the form:

$$\Phi_n = g\tau_c \sum_{n'} K^c_{nn'} \Phi_{n'}, \quad n' \geq 0 \qquad (3)$$

and the dimensionalless quantity g is: $g = \lambda_b(2E_F/3\pi N\tilde{\Omega})$, $\lambda_b$ is the coupling constant for the bulk example. We used the well known McMillan expression : $\lambda_b = <I> \nu / M\tilde{\Omega}^2$ [31], see, *e.g.*, [30], $\nu$ is the density of states, $\nu \propto m p_F$, $p_F$ is the Fermi momentum, $p_F \propto n^{1/3}$, $n$ is the carrier concentration.

The kernel $K^c_{nn'}$ is determined by the expression , which can be obtained from Eq.(1):

$$K^c_{n \neq n'} = \sum_j (f^-_{nn'} + f^+_{nn'})(\tilde{\omega}^2_{n'} + \tilde{\xi}^2_j)^{-1} \qquad (4)$$

$$K^c_{n=n'} \sum_j \{2[1+(2\tilde{\omega}_n)^2]^{-1}(\tilde{\omega}^2_n + \tilde{\xi}^2_j)^{-1} -$$

$$4\sum_{m \neq n} \tilde{\omega}^2_m f^+_{mn} f^-_{mn}(\tilde{\omega}^2_m + \tilde{\xi}^2_j)^{-1}\} \qquad (4')$$

Here $f^\pm_{nn'} = [1+(\tilde{\omega}_n \pm \tilde{\omega}_{n'})]^{-1}$, $\tilde{\omega}_n = \omega_n \tilde{\Omega}^{-1}$, $\tilde{\xi}_s = \xi_s \tilde{\Omega}^{-1}$,

$\xi_s = E_s - \mu$, $E_s$ is the electron energy in normal state, $\mu \equiv \mu(T_c)$ is the chemical potential. Based on Eq. (2), written at $T=T_c$, one can write down the equation allowing us to evaluate $\mu(T_c)$

$$N_e = 2 \sum_j \left[ \frac{(|\xi_j| - \xi_j)}{2|\xi_j|} \right] (X^- + X^+)$$

where $X^{\pm} = 1 + \exp(\pm |\xi_j|/T)$  (5)

The value of $T_c$ can be calculated with use of Eq. (3), or, more specifically, it can be obtained from the matrix equation $\det|1 - K^c_{nn'}|=0$; the matrix $K^c_{nn'}$ is defined by Eqs.(3)-(4'). The calculation for 4X4 matrix provide high accuracy (<1%), cf. Ref. [21]. The value of the energy gap $\varepsilon$ can be calculated from the equation: $\varepsilon = \Delta(-i\varepsilon)$.

As was noted above, the value of $T_c$ can be calculated from Eqs. (3)-(5) and is determined by the values of the parameters $N$, $\tilde{\Omega}$, $E_F$, $\lambda_b$, which are known for each material. The energy levels $E_j$ can be calculated for various models and, in addition, are measured experimentally.

In order to demonstrate the possibility of the transition of the cluster with half-filling into the pairing state let us consider some examples. Note. however, that the determination of the degree of shell filling, as well as "magic" numbers, should be taken with a considerable care. Indeed, as a starting point, the "magic" numbers can be calculated with use of the jellium model (JM, see, *e.g.,* [32]); similar numbers could be obtained by using the potential box model. However, some of these "magic" numbers have not been observed experimentally. The main reason for the deviation is the impact of ionic structure on the spectrum. This impact has been analysed with use of angular photoemission and advanced theoretical methods (see, *e.g.* [15]). As a result, we are dealing with a splitting of the degenerate level obtained by idealized JM and an additional broadening leading to the overlap of the shells.

Let us start with *N=138*, *e.g.* $Zn_{69}$ (each *Zn* atom contains two valance electrons). This is a "magic" number and it is confirmed by theoretical calculations (JM analysis [32] and density functional method) [15]) , and by experimental observations [5,6,15].If we increase a number of atoms, then the next shell with *L=7 (L* is the orbital momentum) becomes partially occupied. The presence of non-occupied states leads to the JT distortion and, correspondingly, to the prolate configuration. The shell with *L=7* contains 28 states. As a result, one can expect that for the cluster with *N=152* (e.g., $Zn_{76}$) one can observe the transition into the pairing state .One can use the following set of the parameters for the $Zn_{76}$ cluster:

$$N = 152, \tilde{\Omega} = 275K, E_F \simeq 1.2 eV \ , \lambda_b = 0.4 \qquad (6)$$

As was noted above, the position of the chemical potential $\mu$ is a very important factor affecting the values of $T_c$ and the energy gap. It is convenient to introduce the dimensionless quantity $\tilde{\mu}$ defined by the relation:

$$\mu = E_H + \tilde{\mu}(E_L - E_H) \qquad (7)$$

$$E_L \equiv E_{LUS}\,; E_H \equiv E_{HOS}.$$

With use of Eqs. (5), (7) we obtain that $\tilde{\mu} \simeq 0$, that is the chemical potential is located near the highest occupied shell. Such location of the chemical potential is very favorable factor for the pairing. Using the values (6), we obtain $T_c \simeq 115K(!)$.

The energy gap value $\varepsilon$ can be determined from Eq.(1) and is a pole of the function $F_s$ with the change $\omega_n \rightarrow -i\varepsilon$. In our case it is of order of $\varepsilon \simeq 3T_c$ and, therefore, $\varepsilon \simeq 345K$ (except in the region near $T_c$). The splitting caused by the dynamic JT effect does not exceed the vibrational spacing. As a result, the gain in energy ($\simeq 2\,\varepsilon$), caused by pairing noticeably exceeds that from the JT splitting.

Let us consider another example ," magic" cluster with $N$=198, e.g., $Al_{66}$. The "magic' nature of this cluster follows from various experimental data [5,6,15}. Moreover, according to [15], the shape of the $Al_{66}$ cluster is very close to being spherical, and this is due to its amorphous structure. Again, an increase in a number of electrons leads to filling of the next shell with $L=8$ [6] and to prolate configuration. Near $N \approx 210$ (for $Al_{70}$ cluster one should expect the next shell to be close to half-filling. Then one can expect the transition into superconducting state. With use of corresponding parameters: $N = 210, \tilde{\Omega} = 350K, E_F = 10^5 K,$

$\lambda_b = 1.4$ (this value of the coupling constant correspond to amorphous $Al$ with $T_{c;b} \approx 6K$ [33]), one can obtain from Eqs.(5),(6) the value of $T_c \approx 175$ K.

The manifestations of pair correlation and corresponding observables are similar to those for the clusters with complete ("magic" clusters) or near complete, energy shells ( see [ 21]). The energy spectrum of an isolated cluster is greatly affected by the pairing and is different above and below $T_c$. More specifically, the value of the excitation energy at low temperatures ($T << T_c$) noticeably exceeds that for $T > T_c$, as follows from the large value of the gap parameter (see above). Moreover, the density of states in the superconducting state, $v_S = v_n[\omega (\omega^2 - \varepsilon^2(T))^{-1/2}]$, strongly depends on temperature dependent order parameter; this dependence is especially strong near $T_c$. In addition, the pair correlation leads to the odd-even effect in the cluster spectrum: the presence of unpaired electron leads to a decrease in the minimum value of the excitation energy for the cluster with odd number of electrons. A similar feature is one of the key observables of the pairing in atomic nuclei. Note also that the cluster-based Josephson tunneling network is capable to transfer a macroscopic superconducting current [34,35].

Observation of the jump in heat capacity for selected clusters [36] (e.g., such a jump was observed at $T \approx 200K$ for $Al_{45}^-$ clusters) was a first indication for superconducting transition for an isolated cluster.

According to recent experimental study [37], the measurements of the ionization potential show that the density of states for some Al clusters, and especially ,for $Al_{66}$ , appears to be temperature dependent. This dependence is in a total agreement with that for the superconducting state (see above) and is caused by the temperature dependence of the gap parameter. According to the analysis [37], the transition occurs at $T_c \approx 120K(!)$. This is the first spectroscopic observation of the phenomenon.

Therefore, the cluster containing a number of delocalized electrons, which is close to the half-filling shell case, can acquire the spherical geometry and this is accompanied by a transition into superconducting state. The nanoclusters $Zn_{76}$, $Al_{70}$ are examples of such systems. Thus, not only "magic" clusters (or clusters with almost filled energy shells, see [21]), but also clusters with near half-filled shells state can display the superconducting state: the electrons in such nano systems can form Cooper pairs at high temperatures.